\documentclass[%
 reprint,
 amsmath,amssymb,
 aps,
 pra,
]{revtex4-2}

\usepackage{graphicx}
\usepackage{dcolumn}
\usepackage{bm}
\usepackage{multirow}
\usepackage{booktabs}
\usepackage[export]{adjustbox}
\usepackage{subfigure}
\usepackage{xcolor}
\usepackage{ulem}

\newcommand{\Slevel}{6\textit{s}$^2$ $^1$\textit{S}$_0$ }
\newcommand{\Dlevel}{5\textit{d}6\textit{p} $^3$\textit{D}$^\text{o}_1$ }

\begin{document}

\title{Direct measurement of isotope shifts in the barium \Slevel - \Dlevel transition}

\author{Jungwoo Choi}\thanks{These authors contributed equally to this work}%
\author{Eunhwi Lee}\thanks{These authors contributed equally to this work}%
\affiliation{SKKU Advanced Institute of Nanotechnology and Department of Nano Science and Technology, Sungkyunkwan University, Suwon, 16419, Korea}

\author{Dahyun Yum}
\altaffiliation{Present address: ID Quantique LTD, SeongNam-si Gyeonggi-do, 13595, Korea}
\affiliation{Department of Physics, Ewha Womans University, Seoul, 03760, Korea}

\author{Kyungwon An}
\affiliation{Department of Physics and Astronomy and Institute of Applied Physics, Seoul National University, Seoul, 08826, Korea}

\author{Junki Kim}
\email{junki.kim.q@skku.edu}
\affiliation{SKKU Advanced Institute of Nanotechnology and Department of Nano Science and Technology, Sungkyunkwan University, Suwon, 16419, Korea \\ 
Department of Nano Engineering, Sungkyunkwan University, Suwon, 16419, Korea}

\date{\today}

\begin{abstract}
    We report the direct measurement of isotope shifts of the barium \textit{\Slevel}--\textit{\Dlevel} 413-nm electric quadrupole transition, which is utilized for efficient barium ion trapping via photoionization using a single coherent light source. 
    The measured isotope shifts relative to $^{138}$Ba are $392.9\pm0.9$ MHz, $178.1\pm0.8$ MHz, $401.4\pm1.2$ MHz, and $124.3\pm1.3$ MHz for isotopes with atomic numbers 137, 136, 135, and 134, respectively.
    We verify the measured isotopes with King plot analysis and compare the result with the formerly known shifts inferred from previous studies on neighboring transitions.
    The results can be used for efficient isotope selective loading of low-abundant barium ions, while careful suppression of line broadening is required for successful isotopic selectivity.
\end{abstract}

\maketitle

\section{Introduction}
Barium ions are among the promising platforms for fundamental science and quantum information technology due to their comparatively long visible wavelength transitions and long-lived metastable D level \cite{dutta_single_2020, dietrich_hyperfine_2010, yum_optical_2017, hannegan_c-band_2021, hannegan_entanglement_2022}.
Recently, odd isotopes of barium ions with non-zero nuclear spin have attracted interest owing to their rich electronic structure and exhibited unmatched state preparation and measurement (SPAM) error \cite{christensen_high-fidelity_2020, an_high_2022}, potential for extended error correction schemes \cite{kang_quantum_2023}, and high-dimensional Hilbert space to store quantum information \cite{low_practical_2020}.
However, scaling the odd-isotope barium ion trap necessitates high isotope-selective ion trapping since their natural abundance is relatively low compared to the most abundant barium-138 isotope.

Among the various ion-loading methods, the resonant photoionization technique has a clear advantage in isotope selectivity \cite{lucas_isotope-selective_2004, kjaergaard_isotope_2000, vrijsen_efficient_2019, gulde_simple_2001}.
In this process, atoms are resonantly excited to an intermediate state with high-isotope selectivity, and an additional laser removes a valence electron to ionize the atoms completely.
So far, three intermediate states have been utilized for resonant photoionization of barium atoms.
The \Slevel--6\textit{s}6\textit{p} $^1$\textit{P}$^\text{o}_1$ transition ($\lambda$ = 553.5 nm) has a strong transition strength for efficient ion loading \cite{yamada_separation_1988, leschhorn_efficient_2012}. However, coherent radiation on the wavelength is only produced by a frequency-doubled laser or dye laser, and the isotope shifts are not sufficiently large for high isotope selectivity \cite{white_isotope-selective_2022}.
The 6\textit{s}6\textit{p} $^3$\textit{P}$_1$ state has large hyperfine splittings that allow efficient isotope selective loading\cite{steele_photoionization_2007, wang_highly_2011}, but the second excitation wavelength falls into the UV region, which reportedly causes photoelectric ionization in an isotope-insensitive manner. 
The \Dlevel intermediate state has high electronic energy, enabling two-photon ionization by a single coherent 413-nm laser source\cite{leschhorn_efficient_2012, rotter_quantum_2008}.
However, isotope shifts of 413-nm transitions and isotope-selective loading using this transition have not yet been reported.

Herein, we report the direct measurement of the isotope shifts in the \Slevel--\Dlevel 413-nm \textit{E}2 transition of atomic barium.
We measured the laser-induced fluorescence spectrum excited by a resonant 413-nm laser source, and the isotope shifts of five naturally abundant barium isotopes were extracted from the spectrum.
The measured shifts are compared with the indirect result of previous spectroscopic studies on neighboring transitions.
Finally, we discuss the feasibility of isotope-selective loading using the \Dlevel intermediate level based on observations.

\section{Experimental setup}
\begin{figure}[h]
    \centering
    \includegraphics[width = 8.6cm]{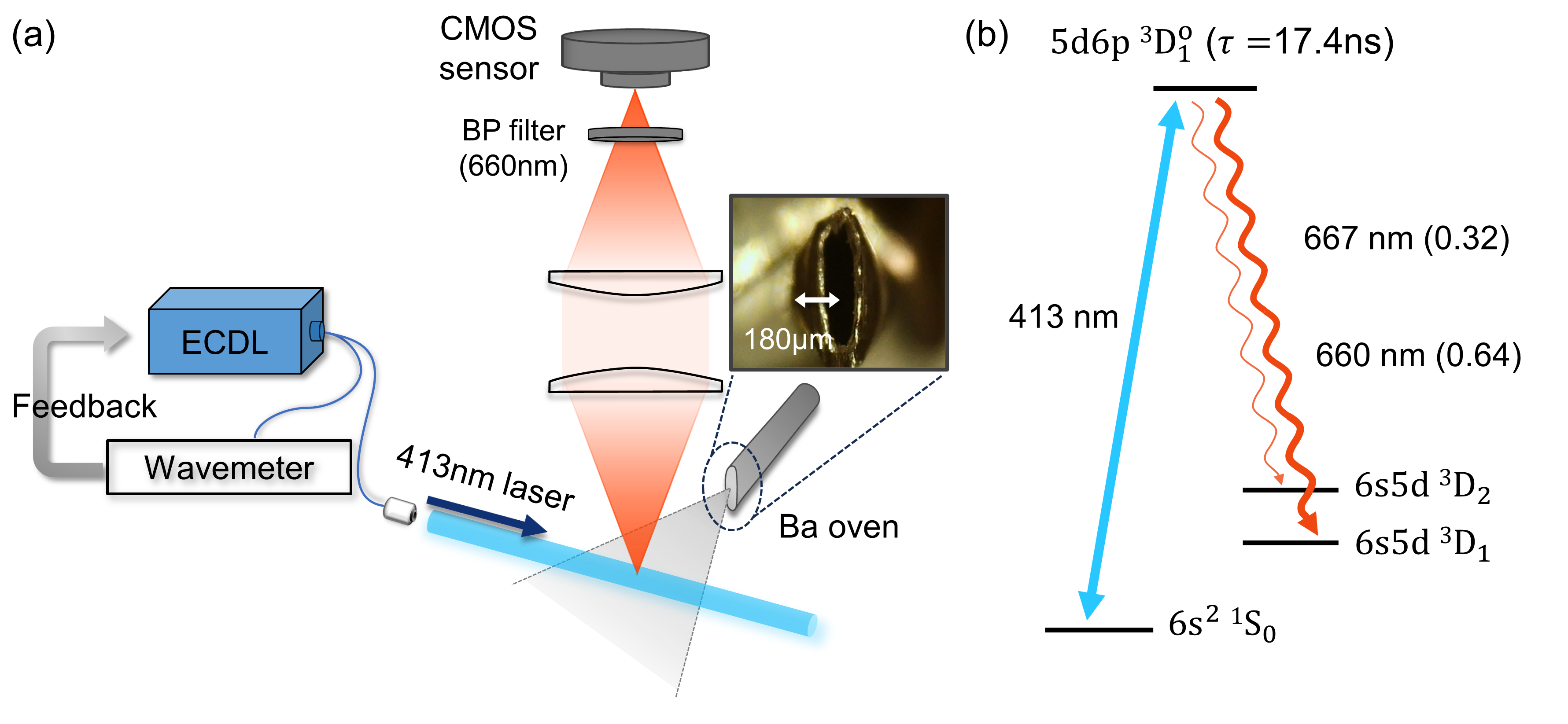}
    \caption{\textbf{(a)} Schematics of experimental setup. The barium atomic beam produced by a thermal oven is excited by a 413-nm laser, and its fluorescence is collected and imaged by the CMOS image sensor. The atomic beam nozzle is compressed along the laser propagation direction to reduce atomic beam divergence in the laser propagation direction. \textbf{(b)} Corresponding electronic structure of barium atoms. The numbers in the parentheses are the branching ratio.}
    \label{fig:schematic}
\end{figure}

Figure \ref{fig:schematic} shows the experimental schematics and related electronic structure of barium. 
The barium atomic beam was produced by resistively heating the Ba metal in a thermal oven, which is placed inside a UHV vacuum chamber with the pressure below $10^{-8}$ Torr. 
The 413-nm laser perpendicular to the atomic beam excites the barium atoms to \Dlevel state, and the excited atoms emit fluorescence photons by spontaneous decay.
The laser beam has a 0.6-mm beam diameter and 60 $\mu$W of beam power and the polarization is perpendicular to the atomic beam.
Atomic fluorescence is collected and imaged by a complementary metal oxide semiconductor (CMOS) image sensor (IMX249, Sony) to investigate the excitation spectrum, while the linear sensor responsivity was verified separately (see Appendix \ref{ap_linear}).
The atomic beam from the oven diverges as it propagates, and the transverse velocity distribution of the region of interest depends on the size of the oven aperture and the distance from the oven nozzle.
To reduce the velocity distribution along the laser beam direction at the interrogation location, we clamped the barium oven aperture and aligned its short axis with a diameter of 180 $\mu$m parallel to the laser beam.
The distance from the oven nozzle to the excitation laser beam is 8.4 mm, and the angular diameter of the clamped nozzle from the region excited by the laser is approximately 2.1 mrad.

The 413-nm excitation laser light is generated by an external cavity diode laser (Toptical DL pro).
The laser frequency was monitored by a high-precision Fizeau-type wavelength meter (HighFinesse WS-8) and stabilized near the target frequency controlled by homemade $\mathrm{PYTHON}$ software \cite{chen_stable_2022}.
While the laser is locked, the frequency error recorded by the wavelength meter has a standard deviation of 0.4 MHz, the same as the wavelength meter resolution.
The wavelength meter was regularly calibrated with a reference 780-nm laser locked to the rubidium-87 \textit{D}2 transition to compensate for the wavelength reading drift. 
The resonant 413-nm light excited the ground-state barium atoms to \Dlevel state with a lifetime of 17.4$\pm$0.5 ns, and the predominant decay channels of the excited state are 6\textit{s}5\textit{d} $^3$\textit{D}$_1$ (64\%) and $^3$\textit{D}$_2$ (32\%) levels, emitting 660-nm and 667-nm photons, respectively. 
An optical 660-nm bandpass filter with a bandwidth of 10 nm before the image sensor blocks ambient lights, and the measured fluorescence is predominantly 660-nm light emitted by atoms.

To acquire the atomic fluorescence spectrum, we scanned the laser frequency by sequentially updating the lock target of wavelength-meter-based frequency stabilization.
Once the laser frequency was stabilized to each target, atomic fluorescence images were achieved from the CMOS sensor. 
The atomic oven was kept on during the process, and the atomic flux from the oven remained constant throughout the experiment.
The entire data acquisition process was automated via network remote procedure calls (RPC) to minimize the system drift.

\section{Result and Discussion}

\begin{figure}[ht]
    \centering
    \includegraphics[width = 8.6cm]{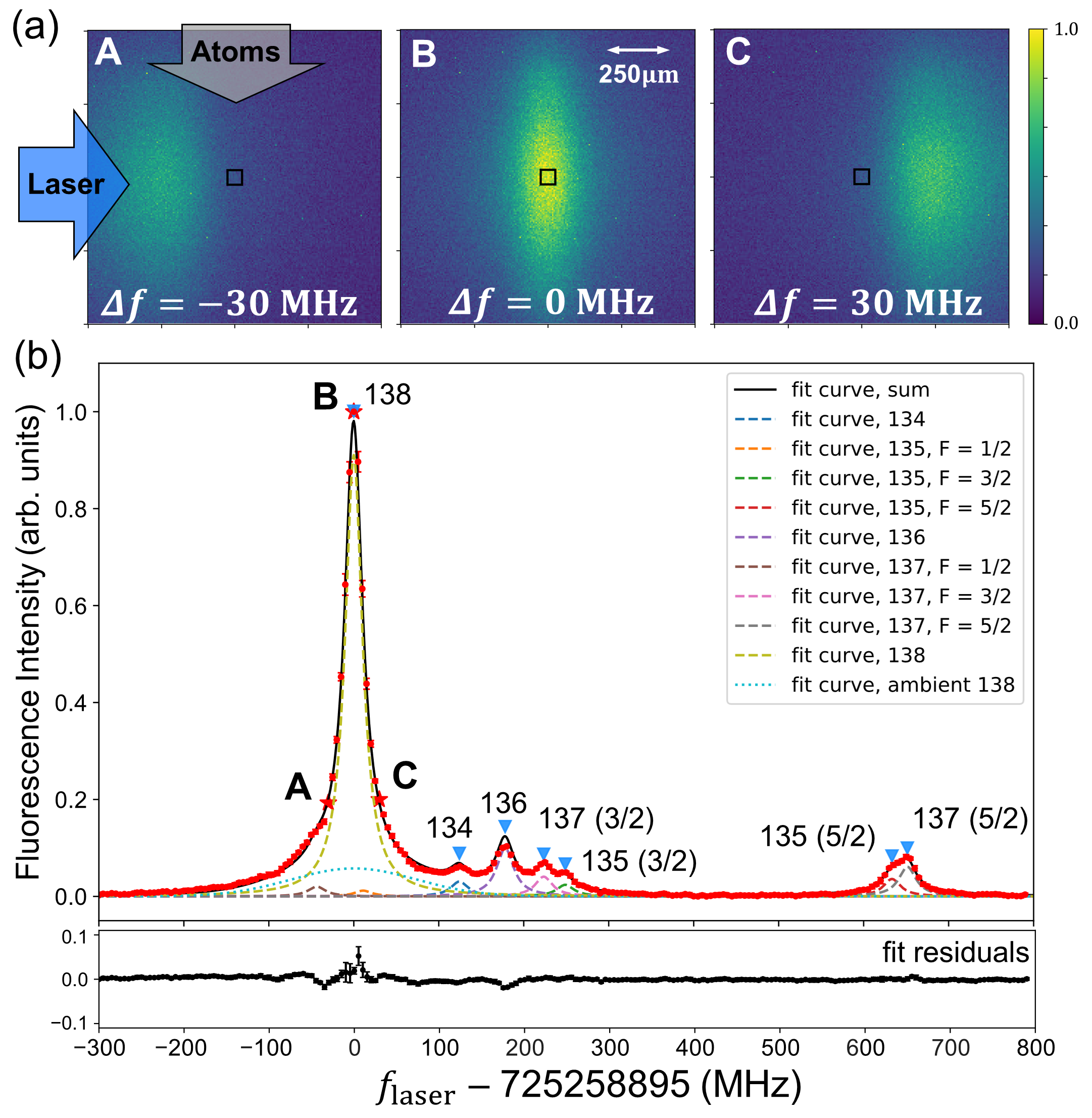}
     \caption{(a) Fluorescence images of barium atomic beam with various laser frequencies ($\Delta f = f_\text{laser} - 7\,255\,258\,895$ MHz). The total fluorescence counts at the perpendicular intersection between atomic beam and excitation laser (black square) are analyzed as spectroscopy signals. The white arrow represents the image scale (250 $\mu$m) where the image aspect ratio is unity. (b) Laser-induced 660-nm fluorescence of atomic barium versus the excitation 413-nm laser frequency. Each data point is an average of ten images, and the error bars represent standard deviations. Solid and dashed lines are the total fit curve and individual fitted peaks, respectively. The bottom plot shows fit residuals. Red stars correspond to the images in (a)}.
    \label{fig:spectrum}
\end{figure}

Figure \ref{fig:spectrum} (a) shows a selection of the atomic fluorescence images obtained. 
As the laser frequency varies, the spatial distribution of atomic fluorescence changes due to the Doppler shift along the laser propagation direction, and the horizontal distribution width becomes the minimum when the light is resonant to the atomic transition.
By comparing the Doppler-broadened spatial distribution of the images, the resonance frequency of the most abundant $^{138}$Ba was found to be 725.258895(5) THz, and the interrogation area where the atomic beam and the excitation laser are perpendicular is also marked [black square in Fig. \ref{fig:spectrum} (a)] (see Appendix \ref{ap_region}).
The laser-induced fluorescence was quantified by summing the fluorescence counts in the region of interest.
Figure \ref{fig:spectrum} (b) shows the excitation spectrum of the \Slevel--\Dlevel transition constructed from fluorescence images collected with different laser frequencies.
Each data point is obtained by averaging ten images, and the step size of the laser frequency was 5 MHz.

In theory, odd isotopes (137 and 135) (\textit{I}=3/2) have three hyperfine sublevels in the excited \Dlevel state, whereas even isotopes (138, 136, and 134) have no hyperfine splitting.
To label each fluorescence peak with the corresponding isotopes and hyperfine levels, we deploy the hyperfine splittings of the \Dlevel state reported in Ref. \cite{grundevik_hyperfine-structure_1983}.
The hyperfine energies can be derived from the Hamiltonian
\begin{equation} \label{eq1}
    H = ah \mathbf{I} \cdot \mathbf{J} + bh \frac{3(\mathbf{I}\cdot\mathbf{J})^2 + 3/2 \mathbf{I}\cdot\mathbf{J} - (\mathbf{I}\cdot\mathbf{I})(\mathbf{J}\cdot\mathbf{J})}{2I(2I-1)J(2J-1)},
\end{equation}
where $h$ is Planck's constant, $a$ is the magnetic dipole interaction constant, $b$ is the electric quadrupole constant, $\mathbf{J}$ is the total angular momentum, and $\mathbf{I}$ is the nuclear spin.
The authors of Ref. \cite{grundevik_hyperfine-structure_1983} reported $a$ and $b$ coefficients of \Dlevel as $a_{135}$ = 155.0$\pm$1.5 MHz, $b_{135}$ = -2.4$\pm$2.0 MHz, $a_{137}$ = 172.9$\pm$0.3 MHz, and $b_{137}$ = -3.5$\pm$0.6 MHz.
By utilizing the derived splitting values, six peaks in the spectrum were successfully matched with their respective isotopes and hyperfine levels, as depicted in Fig. \ref{fig:spectrum} (b).

To determine the precise isotope shifts, we fit the spectrum to the following fit model ($\delta \equiv f - f_{138}$, $f_{138}$ is the barium-138 resonance frequency):
\begin{equation} \label{eq2}
\begin{split}
    I(\delta) &= \sum_{\text{even isotopes}} I_{\text{even}} R_A f(\delta - \delta_A)\\
    & + \sum_{\text{odd isotopes}} \sum_{F} (2F+1)I_{\text{odd}}  R_A f(\delta - \delta_A - \delta_{A,\text{hfs}})
\end{split}    
\end{equation}
with the individual peak function $f(\delta; \gamma, \sigma_{bg}, r_{bg}) = L(\delta; \gamma) + r_{bg} G(\delta;\sigma_{bg})$, the linear combination of Lorentzian distribution $L$ and Gaussian distribution $G$ accounting for the main atomic beam and background ambient atoms, respectively. 
Here, $R_A$ and $\delta_A$ are the natural abundance and isotope shift of the barium isotope $A$ ($\delta_{138}\equiv 0$), $\delta_{A, \text{hfs}}(F)$ is the hyperfine shift of the corresponding hyperfine number $F$, $\gamma$ and $\sigma$ are the Lorentzian and Gaussian widths of the Voigt profile, $\sigma_{bg}$ and $r_{bg}$ are Gaussian widths and relative signal strength of ambient background barium atoms, and $I_\text{even, odd}$ is the amplitude of even and odd isotope peaks due to different polarization response.

 \begin{table}[h!]
        \centering
        \begin{tabular}{cccc}
            \toprule
             Fit parameter&  Fit value & Fit uncertainty & Total uncertainty \\ 
             \hline
             $\delta_{137}$ (MHz)  & 392.9 & 0.6 & 0.9\\
             $\delta_{136}$ (MHz) & 178.1 & 0.4 & 0.8\\
             $\delta_{135}$ (MHz) & 401.4 & 1.0 & 1.2\\
             $\delta_{134}$ (MHz)& 125.3 & 1.2 & 1.3\\
             $f_{138} $ (MHz)& 725258894.7 & 0.2 & -\\
             $\gamma$ (MHz)& 12.3 & 0.2 & - \\
             $\sigma_{bg}$ (MHz) & 73.3 & 2.5 & -\\
             $I_\text{odd}/I_\text{even}$ & $1.19 \times 10^{-2}$  & $3 \times 10^{-4}$ &- \\
             $r_{bg}$ & 0.30 & 0.02 & -\\
            \bottomrule
        \end{tabular}
        \caption{The fit results and total uncertainty}
        \label{tab:fit}
    \end{table}

 \begin{figure}
     \centering
     \includegraphics[width = 8.6cm]{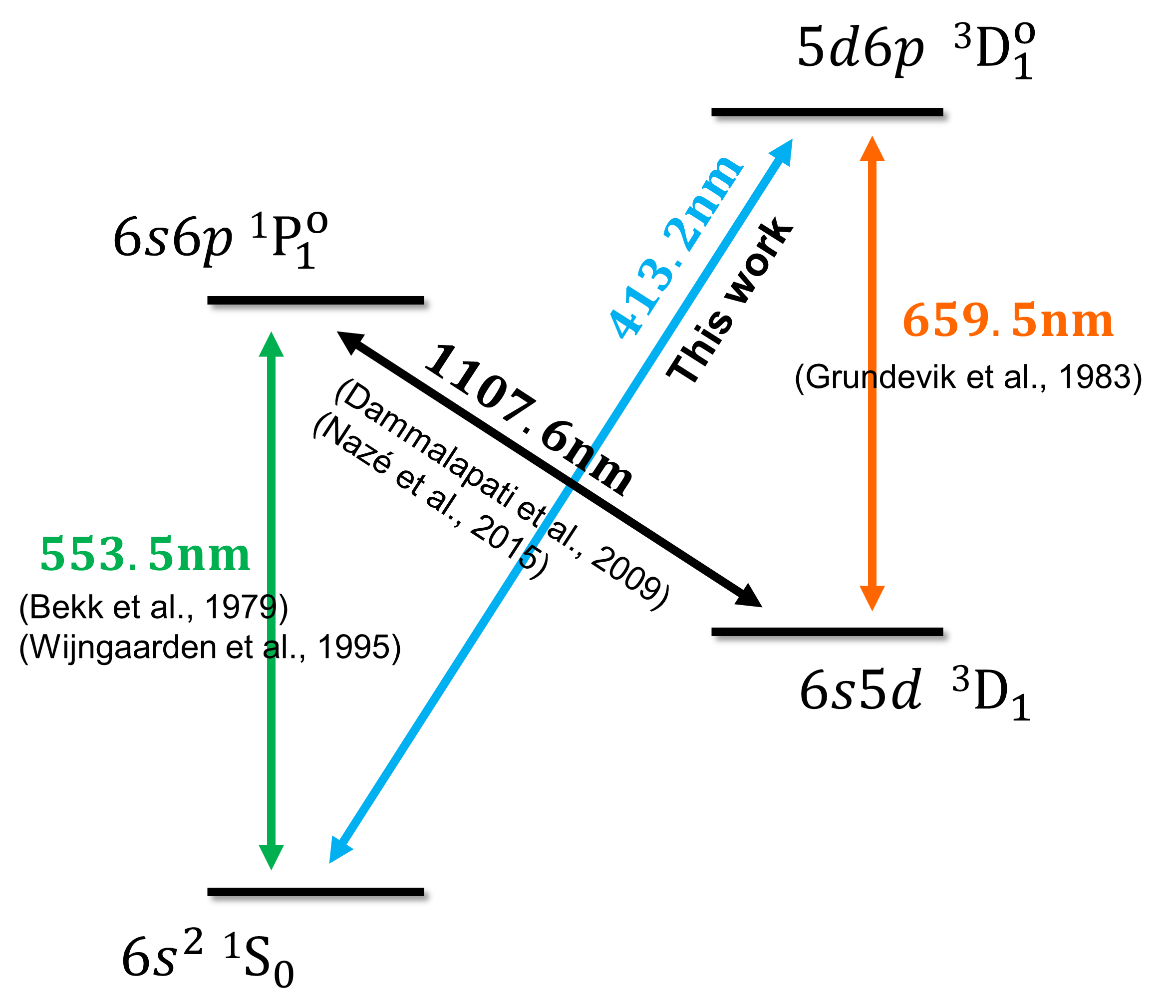}
     \caption{The \Slevel--\Dlevel transition and the neighboring transitions with reported isotope shifts.}
     \label{fig:transition-levels}
 \end{figure}
The model agrees well with the spectrum data and the hyperfine shifts from Ref. \cite{grundevik_hyperfine-structure_1983}, and the isotope shifts of \Slevel--\Dlevel were successfully extracted (Fig. \ref{fig:spectrum} and Table \ref{tab:fit}). 

Other than uncertainty from the fit, our data have uncertainty originated from the frequency measurement via wavelength meter, which is separately measured to be 0.7 MHz (see Appendix \ref{ap_uncertainty}).
Since we are mainly interested in the relative frequency difference between isotope peaks, any common mode errors as laser linewidth and possible angle between atom and laser beam have a negligible effect on the analysis.
By combining the fit uncertainty and measurement uncertainty, we estimate the total uncertainty of our isotope shift measurement.
We find the Lorentzian full-width of fluorescence peaks ($2\gamma = 24.6\pm0.4$ MHz) are relatively large compared to the expected \Dlevel decay rate 9.15 MHz, while the extracted isotope shifts are consistent regardless of peak model (see Appendixes \ref{ap_linewidth} and \ref{ap_peak_model}).

While this work reports the direct observation of isotope shifts in the \Slevel--\Dlevel transition, the isotope shifts of neighboring transitions reported in previous studies may indirectly provide the shift values.
Figure \ref{fig:transition-levels} and Table \ref{table:isotope-shift-summary} summarize the neighboring transitions and their reported isotope shifts.
The \Slevel--6\textit{s}6\textit{p} $^1$\textit{P}$^\text{o}_1$ 553.5-nm transition were investigated in multiple studies \cite{baird_optical_1979, bekk_laserspectroscopic_1979, wijngaarden_hyperfine_1995}, and the reported isotope shifts were consistent with each other.
There were a few experimental studies and a theoretical study on transitions of 6\textit{s}5\textit{d} metastable states \cite{dammalapati_isotope_2009, grundevik_hyperfine-structure_1983, grundevik_hyperfine_1982, naze_theoretical_2015} and by using reported isotope shifts of the transition of 6\textit{s}5\textit{d} $^3$D$_1$--6\textit{s}6\textit{p} $^1$P$_1$ and 6\textit{s}5\textit{d} $^3$D$_1$--\Dlevel, the \Slevel--\Dlevel isotope shifts may be inferred as Table \ref{table:isotope-shift-summary}.
It is noteworthy that the isotope shifts of 6\textit{s}5\textit{d} $^3$D$_1$--6\textit{s}6\textit{p} $^1$P$^\text{o}_1$ 1108 nm from the experimental work \cite{dammalapati_isotope_2009} and the theory work \cite{naze_theoretical_2015} differ from each other and hence infer \Slevel--\Dlevel isotope shifts vary as well.
We find that the even isotope (134 and 136) cases are relatively consistent with each other, exhibiting a mismatch within 15 MHz. 
However, the odd isotope (135 and 137) results show a discrepancy, especially with the experimental work of Dammalapati \textit{et al}. \cite{dammalapati_isotope_2009}.

\begin{table*}[t]
    \centering
    \begin{tabular}{l*5c}
        \toprule
         & \multicolumn{5}{c}{$\Delta \nu^{AA'}$ (\textit{A}=138) (MHz)}\\ 
         $A'$ & 138 & 137 & 136 & 135 & 134 \\
         \hline
         \multicolumn{6}{c}{\Slevel--6\textit{s}6\textit{p} $^1$\textit{P}$_1$ (553.5 nm)}\\ 
         Bekk \textit{et al}. \cite{bekk_laserspectroscopic_1979} & 0.0 & 215.0$\pm$0.7  & 128.9$\pm$0.5  & 260.9$\pm$0.7 & 143.0$\pm$0.5   \\
         Wijngaarden \textit{et al}. \cite{wijngaarden_hyperfine_1995} & 0.0 & 215.2$\pm$0.2  & 128.0$\pm$0.4  & 259.3$\pm$0.2 &  \\ 
         \hline
         \multicolumn{6}{c}{6\textit{s}5\textit{d} $^3$\textit{D}$_1$--6\textit{s}6\textit{p} $^1$\textit{P}$_1$ (1107.6 nm)}\\ 
         Dammalapti \textit{et al}. \cite{dammalapati_isotope_2009} & 0.0 & -114$\pm$4 & 59.3$\pm$0.6  & -39$\pm$4 & 144.1$\pm$1.0   \\
         Nazé \textit{et al}.$^*$ \cite{naze_theoretical_2015} & 0.0 & 67.2  & 79.5  & 132.1 & 139.5 \\ 
         \hline
         \multicolumn{6}{c}{5\textit{d}6\textit{p} $^3$\textit{D}$_1$--6\textit{s}5\textit{d} $^3$\textit{D}$_1$ (659.5 nm)}\\ 
         Grundevik \textit{et al}. \cite{grundevik_hyperfine-structure_1983} & 0.0 & 189.2$\pm$1.0  & 112.9$\pm$1.0  & 228.3$\pm$1.0 & 125.0$\pm$2.0   \\ 
         \hline
         \multicolumn{6}{c}{\Slevel--\Dlevel (413.2 nm) }\\ 
         This work & 0.0 & 392.9$\pm$0.9 & 178.1$\pm$0.8 & 401.4$\pm$1.2 & 124.3$\pm$1.3   \\
         Based on Dammalapti \textit{et al}.\cite{dammalapati_isotope_2009} & 0.0 & 518.2$\pm$4.1  & 182.5$\pm$1.3  & 528.2$\pm$4.2 & 123.9$\pm$2.3 \\
         Based on Nazé \textit{et al}.$^a$ \cite{naze_theoretical_2015} & 0.0 & 337.0$\pm$1.2  & 162.3$\pm$1.1  & 357.1$\pm$1.2 & 128.5$\pm$2.1 \\
        \bottomrule
         \multicolumn{6}{c}{Barium ion (singly charged), 6\textit{s} $^2$S$_{1/2}$--6\textit{p} $^2$P$_{1/2}$ (493.4nm) }\\ 
         Imgram \textit{et al}. \cite{imgram_collinear_2019} & 0.0 & 272.2$\pm$0.2 & 180.2$\pm$0.2 & 350.0$\pm$0.2 & 223.2$\pm$0.2   \\
        \bottomrule         
    \end{tabular}
    \footnotetext{$^*$ Theoretical study}
    \caption{Summary of isotope shifts of the relevant transitions.}
    \label{table:isotope-shift-summary}
\end{table*}

\begin{figure}[b!]
    \centering
    \includegraphics[width = 8.6cm]{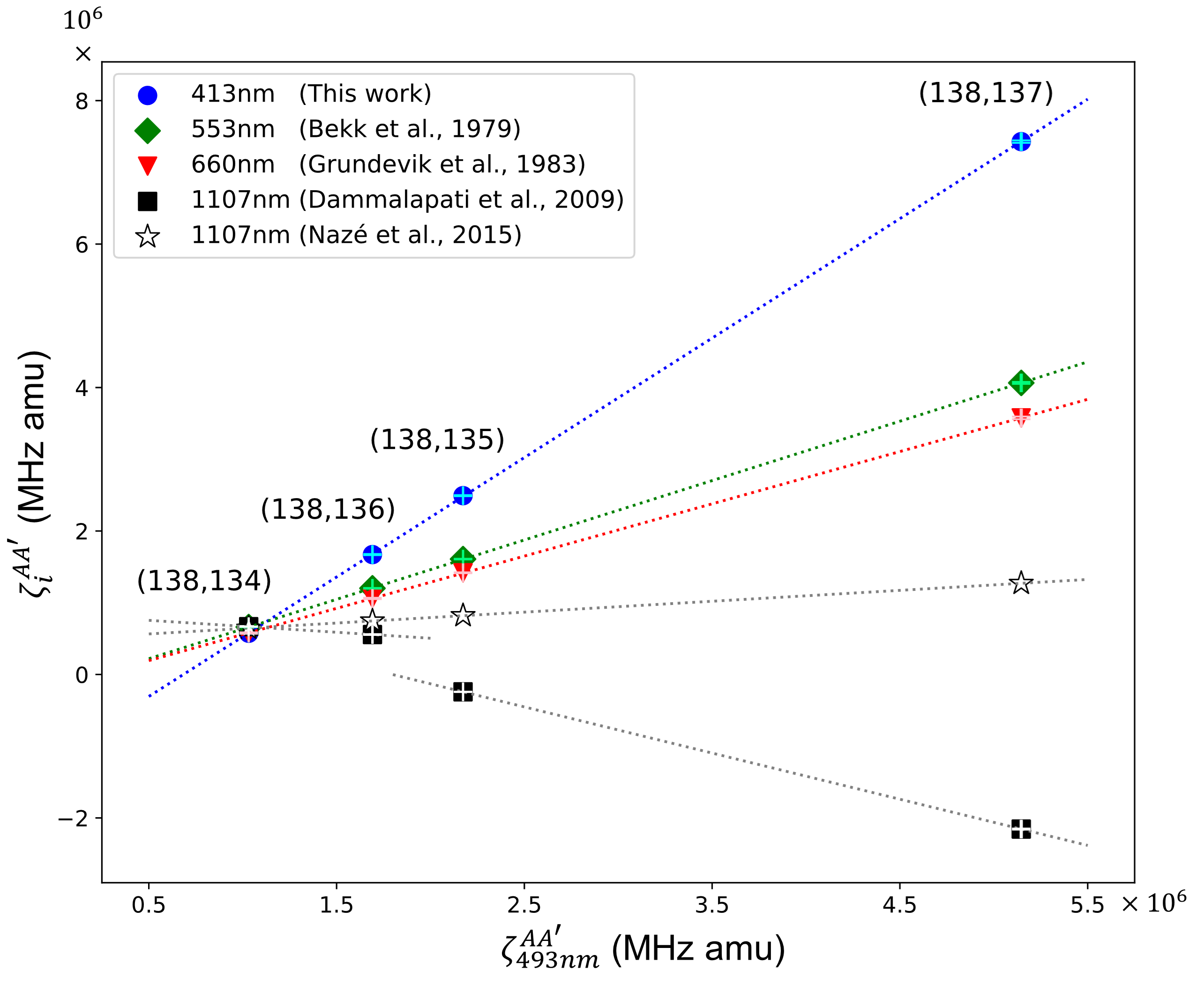}
    \caption{King plot of the relevant transitions in Fig. \ref{fig:transition-levels} with respect to barium ion 6s $^2$S$_{1/2}$ - 6p $^2$P$_{1/2}$ transition.}
    \label{fig:king-plot}
\end{figure}

We compare our result with previous studies using the King plot \cite{king_comments_1963}. 
When we introduce the modified isotope shift $\zeta^{AA'}_i = \Delta \nu^{AA'}_i A'A/(A'-A)$ and plot those shifts in different transitions for all possible pairs $AA’$, the data points are expected to fall on a straight line, which is called the King plot. 
Figure \ref{fig:king-plot} shows the King plot of isotope shifts of our result and previous studies listed in Table \ref{table:isotope-shift-summary} with reference data of the 6\textit{s} $^2$S$_{1/2}$--6\textit{p} $^2$P$_{1/2}$ barium ion transition \cite{imgram_collinear_2019}.
Our experimental data show clear King plot linearity with a slope of 1.63, and other 553.5 nm and 659.5 nm transitions also follow a linear relationship.
For the 1107.8-nm transition, the theoretical work \cite{naze_theoretical_2015} shows linearity, but the work by Dammalapati \textit{et al}. \cite{dammalapati_isotope_2009} shows a mismatch from the expected linearity, as already noted in the study. 
We believe that our study may give a hint for a better understanding of the spectroscopic characteristics of barium atoms.

Lastly, Table \ref{tab:photoionization} shows the 413nm spectrum peaks sorted by their center frequencies, which could be utilized for the resonant isotope-selective photoionization process.
For odd isotopes (135 and 137), the \textit{F}=3/2 state among three hyperfine states has less proximity to
its neighboring peaks, making it a preferred choice for photoionization.
The difference between the peaks of 23.9 MHz is slightly greater than the natural line broadening of 9.15 MHz. 
It could be possible to achieve efficient isotope-selective photoionization by carefully reducing the effects of Doppler broadening and other line broadening factors.

\begingroup
\renewcommand{\arraystretch}{0.6} 
\begin{table}[h]
    \centering
    \begin{tabular}{ccc}
        \toprule
        Peak center (MHz) & Peak label & Peak separation (MHz) \\
        \hline
        \multirow{2}{*}{--43.7} & \multirow{2}{*}{137, \textit{F}=1/2} & \\
          & & \multirow{2}{*}{43.7}           \\
        \multirow{2}{*}{0.0} & \multirow{2}{*}{138} & \\
        &  & \multirow{2}{*}{10.9}           \\
        \multirow{2}{*}{10.9} & \multirow{2}{*}{135, \textit{F}=1/2} & \\
         &  & \multirow{2}{*}{113.3}           \\
        \multirow{2}{*}{124.2} & \multirow{2}{*}{134} & \\
         &  & \multirow{2}{*}{53.9}           \\
        \multirow{2}{*}{178.1} & \multirow{2}{*}{136} & \\
         &  & \multirow{2}{*}{45.4}           \\
        \multirow{2}{*}{\textbf{223.5}} & \multirow{2}{*}{\textbf{137, F=3/2}} & \\
         &  & \multirow{2}{*}{25.3}           \\
        \multirow{2}{*}{\textbf{248.8}} & \multirow{2}{*}{\textbf{135, F=3/2}} & \\
         &  & \multirow{2}{*}{384.5}           \\
        \multirow{2}{*}{633.3} & \multirow{2}{*}{135, \textit{F}=5/2} & \\
         &  & \multirow{2}{*}{18.1}           \\
        \multirow{2}{*}{651.4} & \multirow{2}{*}{137, \textit{F}=5/2} & \\
         &  &            \\
        \bottomrule
    \end{tabular}
    \caption{\Slevel--\Dlevel 413nm transition peak center frequencies. Bold text shows the possible transitions for isotope-selective loading of odd-isotopes}
    \label{tab:photoionization}
\end{table}
\endgroup

In conclusion, the isotope shifts of the barium \Slevel--\Dlevel transition were directly measured from laser-induced fluorescence spectroscopy of the barium atomic beam.
We confirmed the validity of our measurement with King plot analysis and compared the measured isotope shifts with the previous studies on adjacent transition lines.
All nine transition frequencies for different isotopes and hyperfine levels were resolved, and the result suggests careful parameter control is required for isotope-selective loading of low-abundant isotope using 413nm transition.

\begin{acknowledgements}
    This work was supported by the National Research Foundation of Korea (NRF) grant funded by the Korea government(MSIT) (Grants No. 2022R1C1C100375812, No. 2022M3H3A106307412, and No. RS-2023-00302576) and Institute of Information and Communications Technology Planning and Evaluation (IITP) grant funded by the Korean government(MSIT) (Grant No. 2022-0-01040).
\end{acknowledgements}

\clearpage

\appendix

\section{Linearity check of the CMOS image sensor responsivity}\label{ap_linear}
\begin{figure}[ht]
    \centering
    \includegraphics[width = 8.6cm]{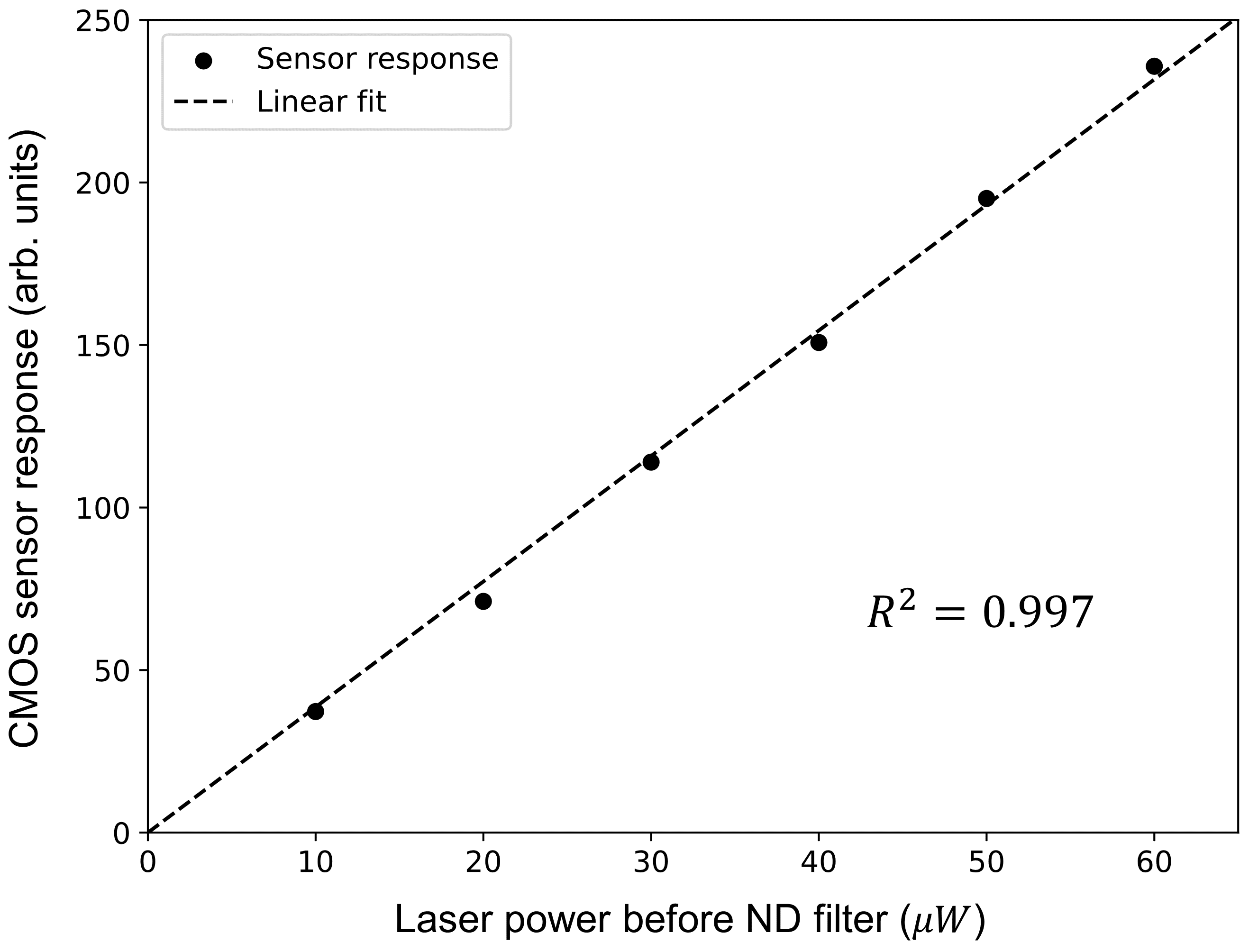}
    \caption{CMOS sensor response to a 650-nm laser diode.}
    \label{fig:cmos-response}
\end{figure}

To verify the linearity of the CMOS sensor responsivity, we observed the sensor response to 650nm laser diode light while varying light intensities.
The image sensor settings, including gain and exposure time, were maintained at the same values as those used in the atomic fluorescence experiment, and the laser intensity was reduced to match the atomic fluorescence intensity by employing a neutral-density (ND) filter with an optical density of 3.8.
The data on Fig. \ref{fig:cmos-response} indicated linear sensor response with the coefficient of determination ($R^2$) for the linear model ($y=ax$) of 0.997.

\section{Selection of the region of interest in the images}\label{ap_region}

\begin{figure}[h]
    \centering
    \includegraphics[width=8.6cm]{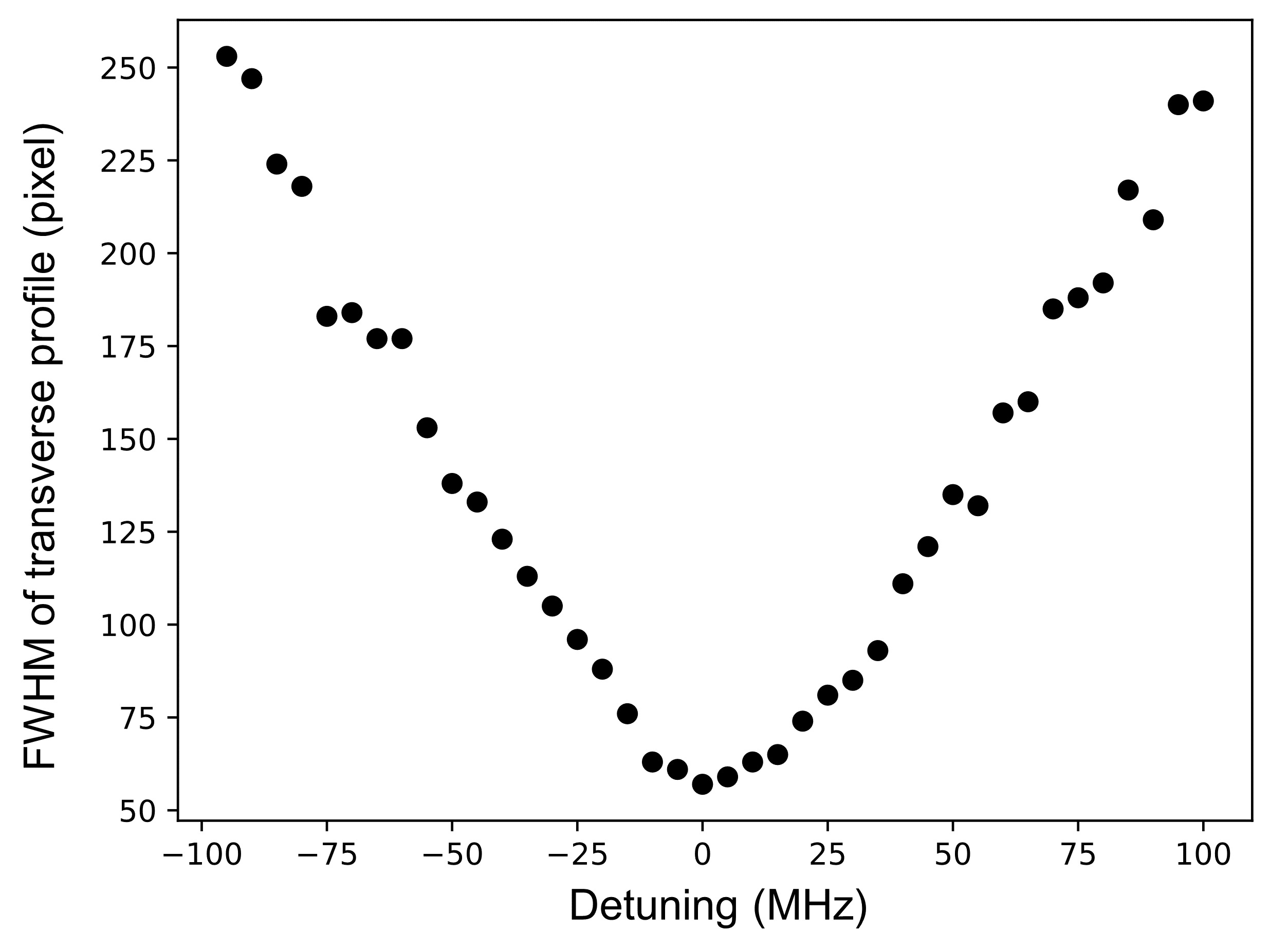}
    \caption{Full-width half-maximum (FWHM) width of transverse profiles of fluorescence images with different laser frequencies. The zero-detuning corresponds to the laser frequency with the minimum distribution width.}
    \label{fig:transverse_width}
\end{figure}

The atomic beam from the oven diverges as it propagates, and the angle between atomic velocity and laser beam varies by interrogation areas.
To achieve a Doppler-free spectrum, we set the image interrogation area such that the atoms traverse perpendicular to the laser beam.
First, we compare the transverse profiles of the acquired fluorescence images along the laser propagation direction.
Due to the high abundance of barium-138, the fluorescence distribution is mainly represented by barium-138 fluorescence.
We observe that when the laser is off-resonant to barium-138, the peak of the transverse profile appears where Doppler shift compensates the detuning between the laser and atomic transition and its width is broader due to Doppler shift by velocity distribution.
Among the images with different laser frequencies, we fin the image with the minimum spatial width and the corresponding frequency is $f=725.258895$THz, suggesting the resonance of barium-138 transition is $f=725.258895(5)$THz (see Fig. \ref{fig:transverse_width}).
Also, we set the region of interest around the fluorescence peak of the resonant image, the square region of 10 $\times$ 10 pixels.

\section{Relative uncertainty of the wavelength meter measurement} \label{ap_uncertainty}

To investigate the relative measurement uncertainty of the wavelength meter, we employ the reference 780-nm laser locked to rubidium \textit{D}2 lines.
The laser is stabilized by saturated-absorption spectroscopy signal with rubidium vapor cell, and the laser frequency can be locked to various hyperfine states of the $^2$P$_{3/2}$ excited state.
The transition frequencies are well known from other studies \cite{arimondo_experimental_1977, ye_hyperfine_1996, banerjee_absolute_2004, steck_rubidium_nodate,steck_rubidium_nodate-1}.
We measure the laser frequency with the wavelength meter while changing the lock target transition.
Table \ref{tab:wlm_relative} shows the laser frequency measurements for various target transitions.
We change the lock set points to six different hyperfine levels of $^{85}$Rb, $^{87}$Rb P$_{3/2}$ excited states and compare the measured laser frequency with the references. 
The measurement result shows consistent agreement within 2 MHz between the measured frequency and the known transition frequency.
We measure the average offset of 0.8 MHz and the standard deviation of data discrepancy is 0.7 MHz while the spanned frequency range is larger than 1.3GHz.
In our main experiment, we mainly focus on the isotope shifts which are frequency differences between two data points and hence common mode error can be ignored.
Therefore, we conclude our relative uncertainty of wavelength meter measurement to be 0.7MHz.

\begin{table}[]
    \centering
    \begin{tabular}{ccccc}
        \toprule
         Isotope and\\ ground state & Excited state & $\Delta f_\text{ref}$(MHz)  & $\Delta f_\text{meas}$(MHz)  \\
         \hline
         \multirow{3}{*}{$^{87}$Rb, $F=2$} & $F'=\text{co}\,1,3$ & --78.5    & --79.1(2)  \\
                                           & $F'=\text{co}\,2,3$ & 0.0      & --0.4(2)   \\
                                           & $F'=3$            & 133.3    & 132.9(2)  \\
         \multirow{3}{*}{$^{85}$Rb, $F=3$} & $F'=\text{co}\,2,4$ & 1167.8   & 1165.9(1) \\
                                           & $F'=\text{co}\,3,4$ & 1199.5   & 1199.5(2) \\
                                           & $F'=4$            & 1259.8   & 1258.1(1) \\
         \bottomrule         
    \end{tabular}
    \caption{Relative frequency uncertainty of the wavelength meter referenced by a laser stabilized to atomic reference. $\Delta f_\text{ref(meas)} = f_\text{ref(meas)} - 384\,227\,981.9 \text{MHz}$, where $f_\text{ref}$ and $f_\text{meas}$ are the transition frequency from references \cite{steck_rubidium_nodate, steck_rubidium_nodate-1} and the measured frequency while the laser is locked to the certain atomic transition.}
    \label{tab:wlm_relative}
\end{table}

\section{linewidths of the fluorescence spectrum}\label{ap_linewidth}

\begin{figure}[h]
    \centering
    \includegraphics[width = 8.6cm]{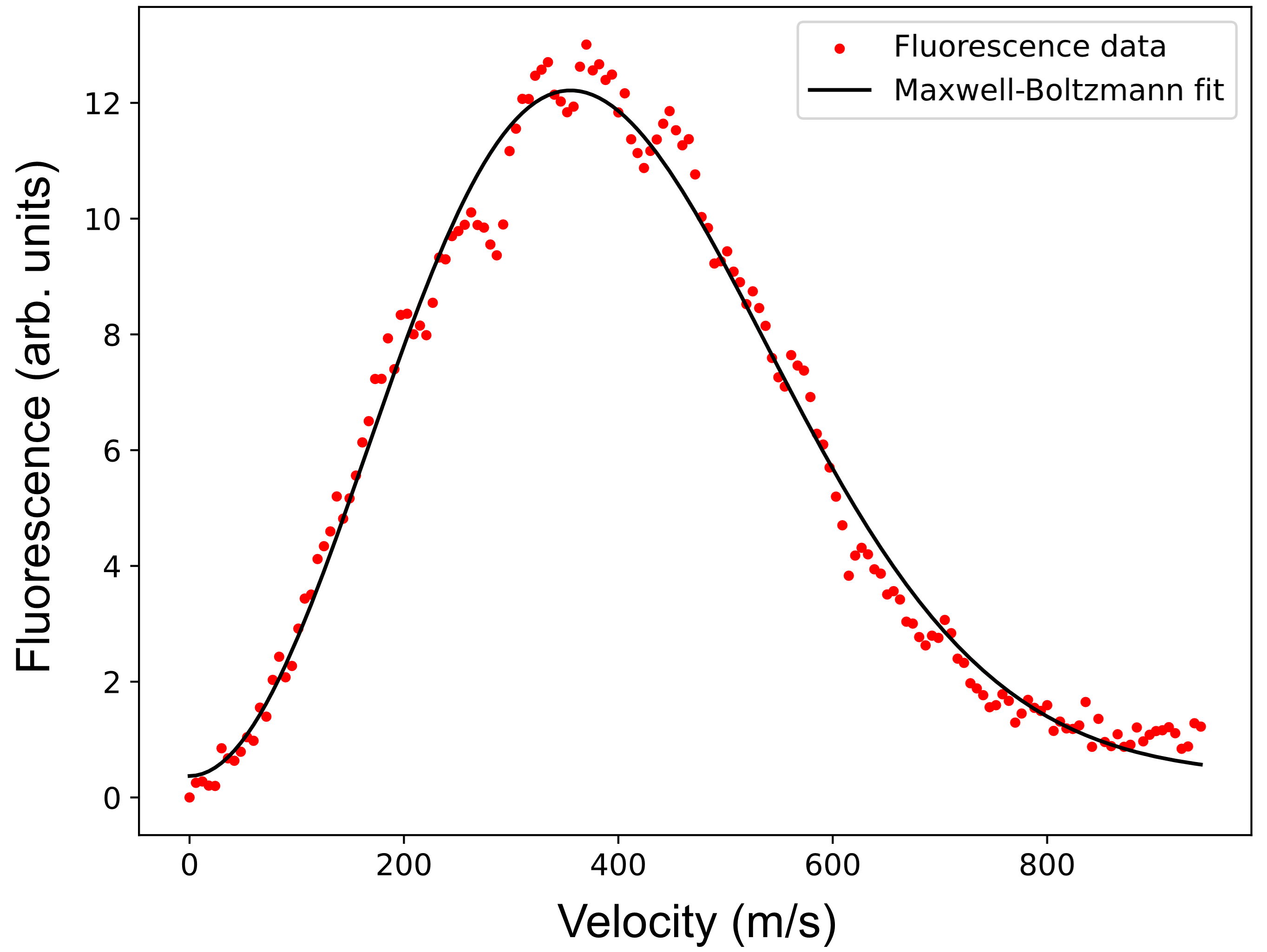}
    \caption{The atomic beam velocity distribution from the Doppler-shifted fluorescence spectrum. }
    \label{fig:velocity}
\end{figure}

\begin{figure}[h]
    \centering
    \includegraphics[width = 8.6cm]{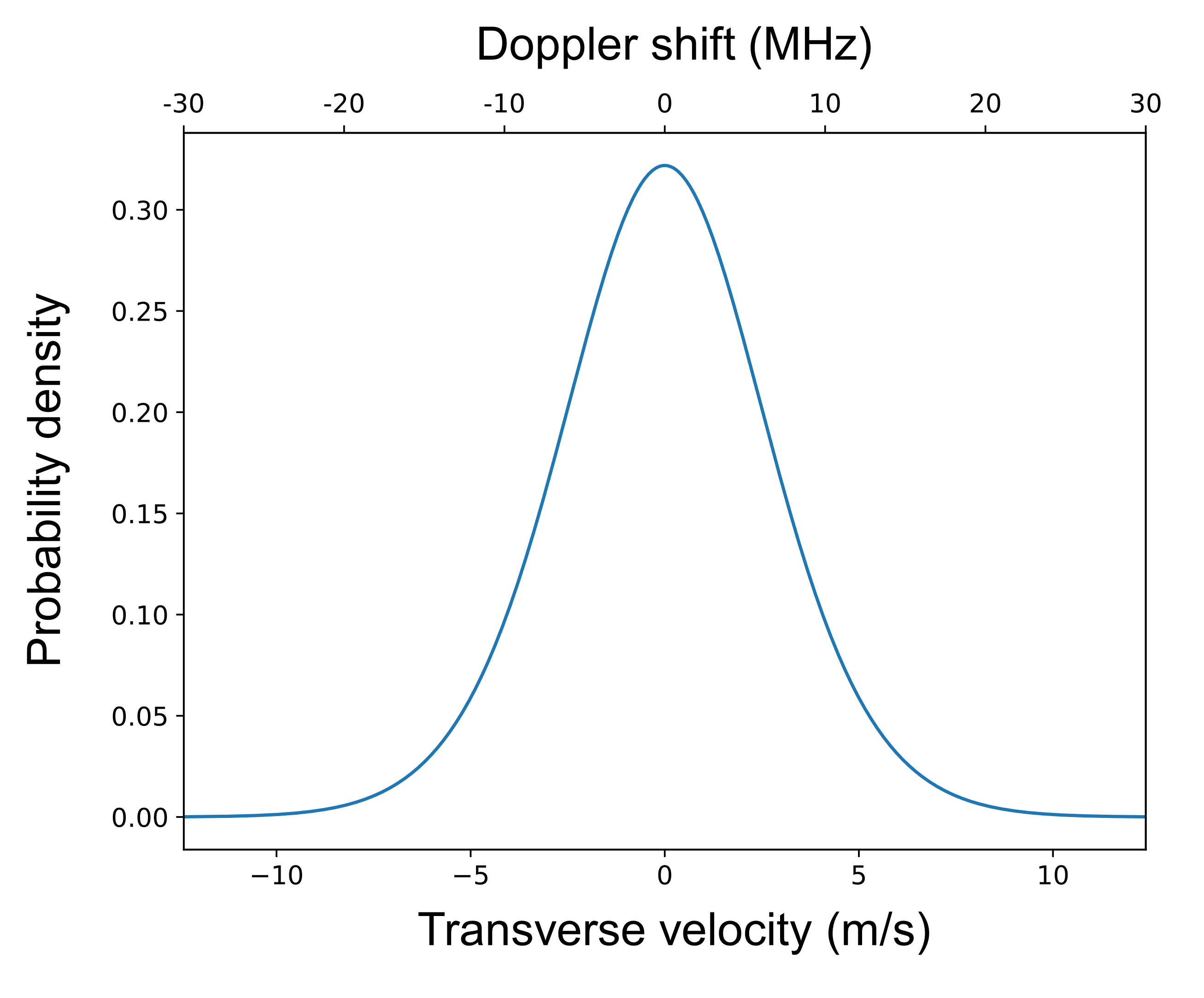}
    \caption{The simulated transverse velocity distribution of a collimated effusive atomic beam. The FWHM of the velocity distribution is 6.15$\pm$0.03 m/s, corresponding to 14.8$\pm$0.1 MHz Doppler broadening width.}
    \label{fig:transvers_velocity}
\end{figure}

To estimate the effect of transverse velocity distribution on the spectrum linewidth, we first measure the velocity distribution of the atomic beam.
We measure the atomic fluorescence spectrum at a point 2.5mm shifted from the normal incident point along the excitation laser direction, where the incident angle between the excitation laser and atoms is 0.35 rad.
By analyzing the Doppler shifts of the spectrum and given geometry between the laser beam and atomic beam, the fit of spectral data to the Boltzmann distribution yielded the root-mean-squared atomic velocity of 435$\pm$2 m/s and temperature of 1050$\pm$8 K (Fig. \ref{fig:velocity}).
From the derived velocity distribution, the transverse velocity profile of a collimated effusive atomic beam could be derived from the model below \cite{greenland_atomic_1985}
\begin{equation}
    \begin{split}
        P_s(v, a_1, a_2) & = |v| \exp{-v^2/2\sigma^2} \\
        & \times \frac{\Gamma(-1/2, v^2/2\sigma^2 S^2) -\Gamma(-1/2, v^2/2\sigma^2 H^2)}
        {4\sqrt{\pi}\sigma^2 (\sqrt{(1+S^2)} - \sqrt{(1+H^2)})}
    \end{split}
\end{equation}
where, $\Gamma(-1/2, x)$ is an incomplete gamma function, $a_1$ is the oven aperture size, $a_2$ is the detection area size, $\sigma$ = kT/m, $S=(a_1 + a_2)/2L,$ and $\, H=(a_2-a_1)/2L$ with the distance $L$ between the oven and detection area.
Figure \ref{fig:transvers_velocity} shows the transverse velocity distribution derived from the velocity distribution, setup geometry, and corresponding Doppler shifts. 
The velocity distribution's full width at half maximum (FWHM) is 6.15$\pm$0.03 m/s, equivalent to Doppler shifts of 14.8$\pm$0.1 MHz. 

\begin{figure}[h]
    \centering
    \includegraphics[width = 8.6cm]{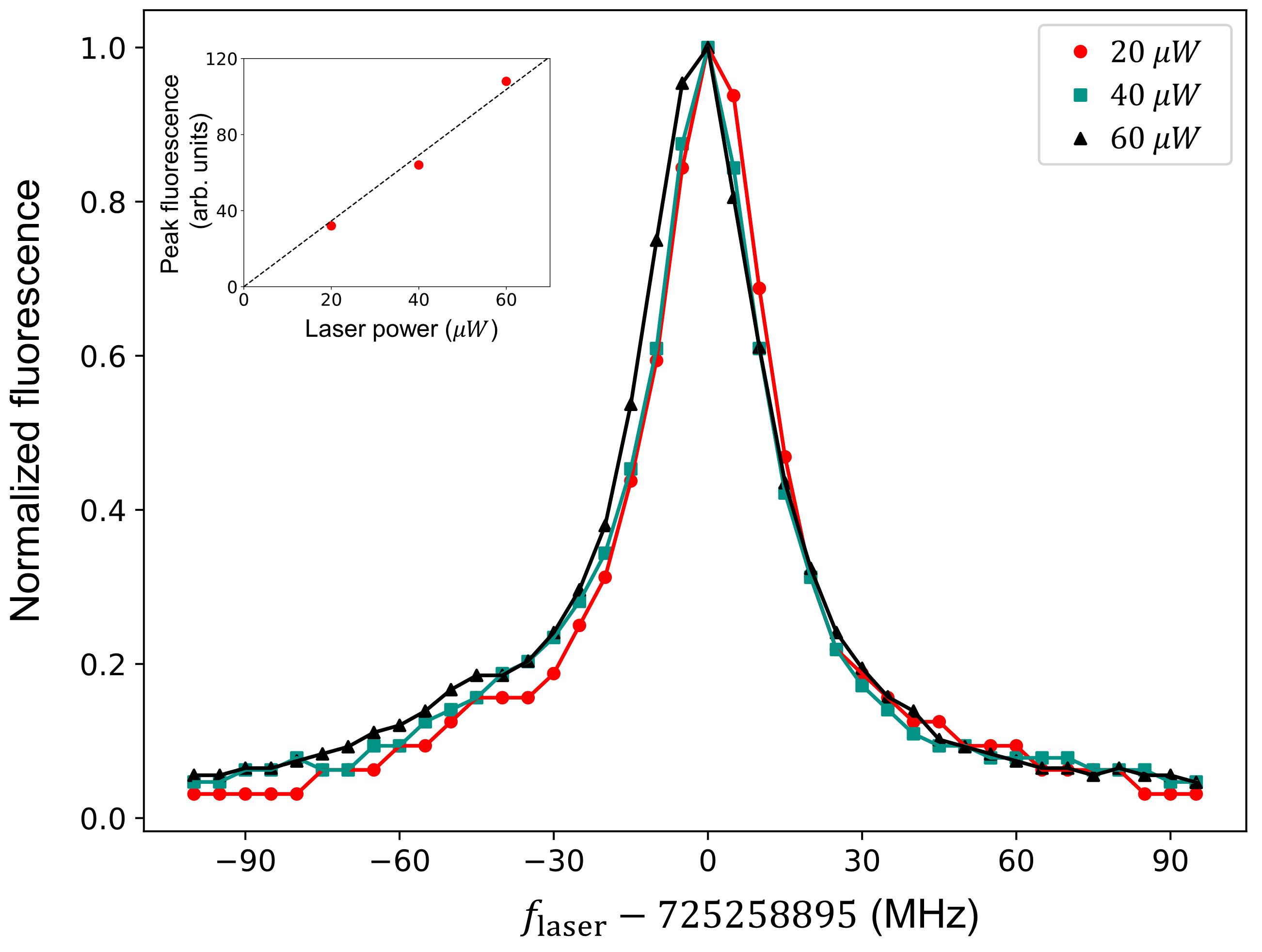}
    \caption{The normalized fluorescence spectra with various 413-nm excitation laser power. The inset shows fluorescence peaks for different excitation laser power, exhibiting linear dependency.}
    \label{fig:power-broadening}
\end{figure}

The spectra with different excitation laser intensities are obtained to confirm the absence of power broadening, and no increase in spectral width is observed (Fig. \ref{fig:power-broadening}).
In addition, the interaction time between the laser field and atoms was about 3 $\mu$s, and the corresponding transit time broadening is negligible.

\clearpage

\section{Comparison of different peak models} \label{ap_peak_model}

Table \ref{table:various models} shows the fit results with various fit function $f(\delta)$ models.
Among the various models, Lorentzian peak with Gaussian background ($L+G$) shows the lowest reduced $\chi^2$ of 1.56, suggesting the model agrees the most with the data.
Although the detailed peak shape differs by the models, the extracted isotope shift values agree within $\pm$ 1 MHz and the choice of peak model does not affect the main result of this work.
With the pure Voigt peak model ($V$), the fitted peak is mostly Lorentzian, while the uncertainty of the Gaussian width is large.
We could also fit the data with the Voigt with the background Gaussian model ($V+G$) while the Lorentizan width is fixed to the natural linewidth ($2\gamma = 9.15$ MHz), although the mean squared error (MSE) is slightly higher than the unfixed case.
The fitted Gaussian FWHM width is $2\sqrt{2 ln2}\sigma = 21.2\pm0.5$ MHz, comparable with the expected Doppler broadening calculated in Appendix \ref{ap_linewidth}.

\begin{table*}
    \centering
    \begin{tabular}{c c c c c }
        \hline
        Fit parameters & $L+G$ & $V$ & $V+G$ & $V+G$ ($\gamma$ fixed)\\        
        \hline
        $\delta_{137}$(MHz) & 392.9 $\pm$ 0.6 & 391.9 $\pm$ 1.2  & 392.9 $\pm$ 0.6  & 392.7 $\pm$ 0.8\\
        $\delta_{136}$(MHz) & 178.1 $\pm$ 0.4 & 176.3 $\pm$ 0.9  & 178.1 $\pm$ 0.4  & 178.3 $\pm$ 0.5\\
        $\delta_{135}$(MHz) & 401.4 $\pm$ 1.0 & 401.3 $\pm$ 2.2  & 401.4 $\pm$ 1.0  & 401.4 $\pm$ 1.4\\
        $\delta_{134}$(MHz) & 124.3 $\pm$ 1.2 & 122.3 $\pm$ 2.3  & 125.3 $\pm$ 1.2  & 126.7 $\pm$ 1.4\\             
        $\gamma$(MHz)       & 12.3 $\pm$ 0.2 & 15.5 $\pm$ 0.5   & 12.3 $\pm$ 0.6   & 4.58 (fixed) \\             
        $\sigma$(MHz)      & N/A              & 0.1 $\pm$ 42.6   & 0.1 $\pm$ 29.2    & 9.0 $\pm$ 0.2\\
        $\sigma_{bg}$(MHz) & 73.3 $\pm$ 2.5   & N/A              & 73.3 $\pm$ 2.7   & 60.3 $\pm$ 1.3\\
        $r_{bg}$       & 0.30 $\pm$ 0.02      & N/A              & 0.30 $\pm$ 0.03  & 0.68 $\pm$ 0.02\\
        \hline
        MSE            & 3.61$\times 10^{-5}$ & 1.35$\times 10^{-4}$ & 3.6 $\times 10^{-5}$ & 6.06$\times 10^{-5}$\\
        Reduced $\chi^2$ & 1.56                & 4.69               & 1.57                & 2.45 \\
        \hline
    \end{tabular}
    \caption{Fit results of various peak models. $L+G$ is the Lorentzian peak with Gaussian background, $V$ is the pure Voigt distribution without background distribution, $V+G$ is Voigt with background Gaussian, and $V+G$ ($\gamma$ fixed) corresponds Voigt with the background Gaussian where Lorentzian width is fixed to \Dlevel decay rate ($2\gamma=9.15$ MHz).}
    \label{table:various models}
\end{table*}

\clearpage


\end{document}